\documentclass[epj]{svjour}

\usepackage{graphics}
\usepackage{color}

\begin{document}
\title{A simple statistical-mechanical interpretation of Onsager
  reciprocal relations and Derjaguin theory of thermo-osmosis}

\author{Oded Farago
  \inst{1,2}
}                     
%
%
\institute{Department of Chemistry, University of Cambridge, Lensfield
  Road, Cambridge CB2 1EW, United Kingdom \and Department of
  Biomedical Engineering, Ben-Gurion University of the Negev, Be'er
  Sheva 85105, Israel}
\date{Received: date / Revised version: date}
%
\abstract{The application of a temperature gradient along a
  fluid-solid interface generates stresses in the fluid causing
  ``thermo-osmotic'' flow. Much of the understanding of this
  phenomenon is based on Derjaguin's work relating thermo-osmotic
  flows to the mechano-caloric effect, namely, the interfacial heat
  flow induced by a pressure gradient. This is done by using Onsager's
  reciprocity relationship for the equivalence of the thermo-osmotic
  and mechano-caloric cross-term transport coefficients.  Both
  Derjaguin theory and Onsager framework for out-of-equilibrium
  systems are formulated in macroscopic thermodynamics terms and lack
  a clear interpretation at the molecular level. Here, we use
  statistical-mechanical tools to derive expressions for the transport
  cross-coefficients and, thereby, to directly demonstrate their
  equality. This is done for two basic models: (i) an incopressible
  continuum solvent containing non-interacting solute particles, and
  (ii) a single-component fluid without thermal expansivity. The
  derivation of the mechano-caloric coefficient appears to be
  remarkably simple, and provides a simple interpretation for the
  connection between interfacial heat and particle fluxes. We use this
  interpretation to consider yet another example, which is an
  electrolyte interacting with a uniformly-charged surface in the
  strong screening (Debye-H\"uckel) regime.}
%
\authorrunning{O. Farago}
\titlerunning{Statistical-mechanical
 approach to Onsager-Derjaguin theory of thermo-osmosis}

\maketitle

\section{Introduction}
\label{sec:intro}

Osmotic flows occur when complex fluids are subject of various
thermodynamic gradients \cite{marbach19}. Examples include
electro-osmosis (flow induced by an gradient of an electric
potential) \cite{wall10}, diffusio-osmosis (solute concentration
gradient) \cite{bacchin19}, and thermo-osmosis (temperature
gradient) \cite{barragan17}. Osmotic flows are interfacial phenomena
arising from stresses induced by the thermodynamics gradients in a
microscopic boundary region, where the properties of the fluid are
influenced by the interactions with the
surface \cite{derjaguinbook}. Osmotic effects are used to manipulate
flows in in microfluidic configurations \cite{stone04}, and in water
treatment and desalination processes \cite{imbrongo16}. Osmosis is
also one of the fundamental transport mechanisms in biological
cells \cite{biologybook}.

Here, we focus on the process of thermo-osmosis which was first
observed in 1907 by Lippmann who measured the flow of water across-a
gelatin membrane separating two reservoirs maintained at different
temperatures \cite{lippmann1907}. Contemporary theoretical
understanding of this phenomenon is largely based on Derjaguin's work
(see chapter 11 in ref.~\cite{derjaguinbook}) who used Onsager’s
linear non-equilibrium thermodynamics (LNET) \cite{onsager1,onsager2}
to relate the thermo-osmotic slip to the fluid-surface interaction
enthalpy (excess enthalpy). Explicitly, Onsager's phenomenological
theory expresses the heat and material fluxes of a complex fluid as
linear combinations of the applied temperature and pressure
gradients. From general considerations (to be reviewed below), the
cross-terms of the transport matrix, namely the thermo-osmotic
coefficient describing surface-induced flow under a temperature
gradient $\vec{\nabla}T$ and the mechno-caloric coefficient describing
heat flux due to a pressure gradient $\vec{\nabla}P$, must be equal to
each other.

Application of the theory requires some dynamical model for the
relevant variables. In~\cite{derjaguinbook}, Derjaguin considered a
simple liquid and calculated the slip velocity from the interfacial
thermo-osmotic flow using the Navier-Stokes (NS) hydrodynamic
equation. A simple analytical expression for the slip velocity can be
obtained by assuming no-slip boundary conditions at the surface, and
by ignoring the variations in the viscosity of the fluid due to the
interactions with the surface. Further assuming that the boundary
layer is much smaller than the size of the ``flow chamber'', then a
linear approximation can be used for the flow field, leading to
Derjaguin's formula for the slip velocity \cite{derjaguinbook}
\begin{equation}
v_s=-\left[\frac{1}{\eta}\int_0^{\infty} z\Delta
  h(z)dz\right]\frac{\vec{\nabla}T}{T},
\label{eq:slipvel}
\end{equation}
where $z$ is the perpendicular distance from the surface, $\Delta
h(z)$ is the excess enthalpy density, and $\eta$ is the fluid
viscosity. Equation~(\ref{eq:slipvel}) has been recently used to
measure the thermo-osmotic slip experimentally \cite{bergulla16} and
in computer simulations \cite{ganti17}.

Onsager's reciprocity principle is derived based on the general
assumption of microscopic reversibility and does not involve
hydrodynamic considerations \cite{grootmazurbook}. The theory is not
limited to thermo-osmosis but applies to a large class of transport
phenomena, and it does not address examples of specific molecular
systems. Here we present a complementary statistical-mechanical
approach and consider thermo-osmosis in two fundamental models -\\ (i)
solvent-solute mixtures, and (ii) a single-component fluid. In each
model we derive expressions that explicitly relate the thermo-osmotic
and mechano-caloric coefficients to equilibrium properties of the
model systems. As expected, the derived expressions are identical to
each other, which provides a direct demonstration of Onsager's
principle. In the former example of a solvent-solute mixture, the
solvent is treated as an incompressible stuctureless fluid, which
facilitates the calculation of the interfacial stress
\cite{wurger10}. It is, however, clear that the excess enthalpy
originates from the solute-surface rather than the solvent-surface
interactions. Therefore, in the calculation of the transport
cross-coefficients, the solute flux must be considered, which is not
captured by the NS hydrodynamic equation but rather by a Smoluchowski
diffusion equation.

Related to the above last point, we wish to emphasize that the focus
on this paper is on the equality of Onsager's transport
cross-coefficients, which is the basis for Derjaguin's theory of
thermo-osmosis. Accordingly, the paper is organized as follows: In
section~\ref{sec:onsderj} we provide a simple derivation of this
fundamental result, which is based on general thermodynamic
considerations. Then we use statistical-mechanical tools to derive
explicitly expressions for the transport cross-coefficients, thereby
demonstrating their equality, in two basic models - the solvent-solute
mixtures (section~\ref{sec:solsol}), and a single-component fluid
(section~\ref{sec:1component}). We summarize the main results in
section~\ref{sec:summary}, where we briefly discuss another example of
an electrolyte solution. The statistical-mechanical derivations
involve some basic assumptions, most importantly, the concept of local
thermal equilibrium. At the molecular level, this means that when the
particles move across a temperature gradient, their velocity
distribution quickly adopts to the local temperature and, therefore,
their dynamics must be non-inertial. Following from this assumption is
the fact that the temporal evolution of the mass and surface heat
energy densities can be described by corresponding diffusion
equations, and that in a steady state, the divergences of the relevant
currents vanish. The very same basic assumptions have been recently
used by Anzini et al.~to develop a field-theoretical approach for
thermo-osmosis \cite{anzini19}. However, the focus on that work was
not on the Onsager's cross-coefficients, but on the fluid slip
velocity (\ref{eq:slipvel}), which is a different part of Derjaguin's
theory that requires taking into account hydrodynamic considerations
(liquids) or kinetic theories (gases). Our approach, in contrast, is
purely statistical-mechanics and, therefore, we only briefly discuss
the hydrodynamic aspects of Derjaguin's theory in
appendix~\ref{sec:hydro}. A key point in the statistical-mechanical
approach presented herein is the assumption that Einstein's relation
between the mobility, temperature, and diffusion coefficient, holds at
the local level, which is consistent with the local equilibrium
picture. In this approach, the hydrodynamic or kinetic behavior of a
specific system are implicitly accounted by the local mobility of the
transported particles.

\section{Onsager-Derjaguin formulation}
\label{sec:onsderj}  

Consider a thin slab of width $dx$ and cross-sectional area $L_y\times
L_z$, which is sufficiently large to be treated as a macroscopic
thermodynamic system. Along the $x$ direction, the system is coupled
to two thermodynamic reservoirs with slightly different temperatures
($T$ and $T+\Delta T$) and chemical potentials ($\mu$ and $\mu+\Delta
\mu$). The temperature and chemical potential gradients cause
steady-state heat, $J_Q$, and particle, $J_N$, currents to flow
throughout the slab. These currents from one reservoir to the other
generate entropy, and from the fundamental thermodynamic equation for
the differential change in the entropy, $dS=dU/T+dV(P/T)-dN(\mu/T)$,
it follows that the rate of net entropy production, $\dot{S}$, is equal to
\begin{equation}
  \dot{S}=J_Q\Delta\left(\frac{1}{T}\right)
  -J_N\Delta\left(\frac{\mu}{T}\right).
  \label{eq:sproduction}
\end{equation}
We note that the three intensive thermodynamic variables ($T$, $P$,
and $\mu$) are not mutually independent. Thus, $\mu=\mu(T,P)$, and in
terms of the temperature and pressure gradient $\Delta P$,
Eq.~(\ref{eq:sproduction}) reads
\begin{equation}
  \dot{S}=J_Q\Delta\left(\frac{1}{T}\right)
  \!-\!J_N\!\left[\frac{\partial}{\partial T}
    \left(\frac{\mu}{T}\right)_P\Delta T+
   \frac{\partial}{\partial P}\left(\frac{\mu}{T}\right)_T\Delta P\right].
  \label{eq:sproduction2}
\end{equation}
From the Gibbs-Helmholtz relation we have that\\
$\partial(\mu/T)/\partial T|_P=-H/NT^2$, where $H$ is the
enthalpy. Also, $\mu=G/N$, where $G$ is the Gibbs free energy and,
therefore, $\partial(\mu/T)/\partial P|_T=(V/NT)$. Using these
relations in Eq.~(\ref{eq:sproduction2}), leads to the following form
\begin{equation}
  \dot{S}=\left(-J_Q+J_H\right)\frac{\Delta T}{T^2}+J_V\frac{\Delta P}{T},
  \label{eq:sproduction3}
\end{equation}
where $J_H=J_NH/N$ is the enthalpy current, and $J_V=J_NV/N=J_N/\rho$
is the volume material current. Importantly, while the current $J_Q$
represents the flow of internal energy arising from the intramolecular
forces between the constituent particles, the current $J_H$ descries
the total enthalpy flow. Thus, the difference between them is the
excess enthalpy current associated with the interactions of the
particles with the surfaces that bound the system. Denoting the excess
enthalpy current by $J_{H_s}=J_H-J_Q$, we arrive at Derjaguin's
formula for the rate of entropy production, $\dot{\cal Q}\equiv
T\dot{S}$ \cite{derjaguinbook}
\begin{equation}
  \dot{\cal
    Q}=\left[J_{H_s}\frac{T^{\prime}}{T}+J_VP^{\prime}\right]dx.
  \label{eq:sproduction4}
\end{equation}
For consistency with forthcoming calculations, let us assume that the
particles interact only with the surfaces that bound the system in the
$z$ direction, but not in the $y$ direction. Rather than dealing with
extensive thermodynamic quantities, we can divide
Eq.~(\ref{eq:sproduction4}) by $VL_y=dxL_y^2L_z$ and consider the
entropy density (per unit volume) production, per unit length in the
$y$ direction
\begin{equation}
  \frac{\dot{\cal Q}}{dxL_y^2L_z}\equiv\dot{\cal
    R}=j_{h}\frac{T^{\prime}}{T}+j_nP^{\prime},
  \label{eq:sproduction5}
\end{equation}
where $j_h$ and $j_n$ denote, respectively, the excess enthalpy density and
particle fluxes, per unit length.

The Onsager reciprocity relation can be now obtained following a
simple route \cite{berreta04}.  We write the phenomenological linear
relation between the fluxes $j_{h}$ and $j_n$, and the thermodynamic
forces, $T^{\prime}/T$ and $P^{\prime}$
\begin{equation}
  \left[ \begin{array}{c} j_{h} \\ \\ j_n \end{array} \right]
      = \left[\begin{array}{cc}
        L_{hh} & L_{hn} \\ \\ L_{nh}  & L_{nn} \end{array}\right]
      \left[ \begin{array}{c} \frac{T^{\prime}}{T} \\ \\ P^{\prime}
        \end{array} \right]
\label{eq:linmatrix}
\end{equation}
Then, from Eqs.~(\ref{eq:sproduction5}) and (\ref{eq:linmatrix}) we
conclude that
\begin{equation}
  L_{hn}=\frac{\partial j_h}{\partial
    P^{\prime}}=\frac{\partial^2\dot{\cal R}}{\partial
    P^{\prime}\partial (T^{\prime}/T)}=\frac{\partial j_n}{\partial
    (T^{\prime}/T)}=L_{nh},
  \label{eq:onsagerrel}
\end{equation}
which is the Onsager relation relevant to thermo-osmosis. Note that in
the notation of fluxes per unit length, the cross-coefficient
$L_{hn}=L_{nh}$ have units of frequency (inverse time).

\section{Solvent-solute systems}
\label{sec:solsol}

Osmotic flows have been frequently studied within the context of
liquids containing solute molecules, e.g., polymer suspensions or
ionic solutions. In many theoretical studies the solvent is treated as
an incompressible medium, while the solute is often assumed to be at
low concentration and considered as a gas of particles interacting
with the surface but not with each other. Much of this framework has
been put forward by Anderson et al.~in their 1982 paper on
diffusophoresis of colloids in non-electrolytes \cite{anderson82}. The
largely accepted picture is that the solute interactions with the
surface of the colloid lead to an osmotic pressure difference in the
solvent due to the preferential (or dispreferential) adsorption of
solute particles. Application of a temperature (or a chemical
potential) gradient would then generate an asymmetric distribution of
solute around the colloid and, thus, a pressure gradient which, in
turn, drives a tangential flow of the solvent. Treating the solvent as
an incompressible liquid with constant viscosity allows one to
calculate the thermo-osmotic force exerted on the surface via the NS
equation (see below). In this picture, however, the solvent has no
influence on the non-equilibrium thermodynamics of the system and,
specifically, on the Onsager transport coefficients. These must be
calculated from the fluxes of the solute particles rather than the
flow velocity of the solvent. We now derive expressions for the cross
coefficients $L_{hn}$ and $L_{nh}$, relating them to equilibrium
properties of the solute. The derived expressions demonstrate that,
indeed, $L_{hn}=L_{nh}$.

We consider a semi-infinite system which is bound at $z=0$ by a
surface, with which the solute particles interact via a short-range
interaction potential $u(z)$. We further assume that the concentration
of solute particles, $c$, is low and, thus, ignore their interactions
with each other. Denoting by $c_0$ the bulk concentration at
$z\rightarrow\infty$, the equilibrium concentration at constant
temperature $T_0$ is given by
\begin{equation}
  c(z)=c_0e^{-u(z)/k_BT_0},
  \label{eq:pmf2}
\end{equation}
where $k_B$ is Boltzmann's constant. We now wish to calculate the
mechano-caloric and thermo-osmotic Onsager coefficients, and
demonstrate their equivalence.  In the following derivations, we will
use two commonly made approximations: The first one, which has also
been used in Anderson's framework ref. \cite{anderson82}, is that the
P\'eclet number is small. This means that the term associated with the
solvent velocity field relative to the solute particles can be ignored
in the expression for the solute current. The second one is the local
thermal equilibrium (LTE) approximation that underlies the classical
thermodynamic approach to the problem. In this approximation, the
concentration at $(x,z)$ is given by Eq.~(\ref{eq:pmf2}) with
$T=T(x)$: $c(x,z)=c_0(T(x))\exp[-u(z)/k_BT(x)]$. This local
equilibrium distribution is not affected by the application of a
pressure gradient that causes the particles to flow, but at a rate
which is assumed to be sufficiently low such that they quickly adopt
to the local temperature.

We begin with the calculation of the mechano-caloric coefficient by
assuming constant temperature $T(x)=T_0$ and a small uniform pressure
gradient along the $x$ direction: $P(x)=P_0+P^{\prime}x$.  The
application of a small external pressure gradient,
$P(x)=P_0+P^{\prime}x$ generates a constant flux of solute particles
which is equal to $\vec{j}_n=-\mu_0P^{\prime}\hat{x}$, where $\mu_0$
is the mobility of the solute particles. The excess heat flow is
associated with the interfacial interaction energy that the moving
particles carry. Thus, the excess heat (enthalpy) density flux at
distance $z$ from the surface is then given by
$\vec{j}_h(z)=u(z)c(z)\vec{j}_n=-u(z)c(z)\mu_0P^{\prime}\hat{x}$. The
total heat flux per unit length is obtained by integration
$\vec{j}_{h}=\int_0^{\infty}\vec{j}_h(z)dz$, and by comparison with
Eq.~(\ref{eq:linmatrix}), we conclude that
\begin{equation}
  L_{hn}=- \mu_0c_0\int_0^{\infty} u(z)e^{-u(z)/k_BT_0}dz.
\label{eq:lhp2}
\end{equation}

The calculation of the thermo-osmotic coefficient $L_{nh}$ is more
complicated. It begins with the assumptions that the system is driven
out of equilibrium by the application of a small temperature gradient,
$T(x)=T_0+T^{\prime}x$, while the pressure is maintained uniform. The
solute flux per unit volume is now given by
\begin{equation}
  \vec{j}_n=-\mu_0\left\{\vec{\nabla}\left[k_BT(x)c(x,z)\right]
  +c(x,z)\vec{\nabla}u\right\},
  \label{eq:solflux}
\end{equation}
where the first term takes into account the fact that the diffusion
coefficient $D(x)=\mu_0k_BT(x)$ depends on the coordinate $x$. The
steady-state flux can be found by solving the equation
\begin{equation}
  \vec{\nabla}\cdot\vec{j_n}=0,
  \label{eq:eqss}
\end{equation}
subject to the boundary condition at $z\rightarrow \infty$ that\\
$c(x)T(x)=c_0T_0$. This condition expresses the fact that the bulk
osmotic pressure of the solute, which is treated as an ideal gas of
non-interacting particles, must be
uniform \cite{fayolle08,farago19}. Thus, far away from the surface,
$c(x,z\rightarrow\infty)=c_0T_0/T(x)$ which, to linear order in
$T^{\prime}$, can be also written as
\begin{equation}
  c(x,z\rightarrow\infty)\simeq
  c_0\left[1-\frac{T^{\prime}x}{T_0}\right].
  \label{eq:bc}
\end{equation}
The non-uniform bulk concentration is a manifestation of the Soret
effect (see extensive review in~\cite{piazza08}). Typically, the
dominant contribution to this effect arises from the inter-molecular
forces. These generate effective mechanical thermophoretic forces on
the molecules that distort their distribution either in the direction
or opposite to the temperature gradient. In this section we consider
an ideal gas of solute particles and, therefore, the only contribution
to the Soret effect is the, so called, ideal gas term arising from the
random thermal collisions whose strength grows with the local
temperature (see discussion in~\cite{fayolle08,farago19}). This is the
reason for the absence of a thermophoretic force term in
Eq.~(\ref{eq:solflux}).

Obviously, in the absence of a temperature gradient ($T(x)=T_0$) the
solution of Eq.~(\ref{eq:eqss}) is the equilibrium distribution
(\ref{eq:pmf2}). Under the influence of a small temperature gradient,
the solution assumes the form: $c(x,z)=c_0e^{-u(z)/k_BT_0}+c_1(x,z)$,
where the perturbation, $c_1$, is, to leading order, linear in the
temperature gradient: $c_1\sim T^{\prime}$. Substituting this form of
the solution in the steady-state Eq.~(\ref{eq:eqss}) and keeping only
linear terms in $T^{\prime}$, leads to the following equation for
$c_1(x,z)$:
\begin{eqnarray}
&k_BT_0\nabla^2 c_1+\frac{\partial c_1}{\partial
  z}u^{\prime}(z)+c_1u^{\prime\prime}(z) \label{eq:eqc1}\\
&=-k_BT^{\prime}xc_0\left[-\frac{u^{\prime\prime}(z)}{k_BT_0}
    +\left(\frac{u^{\prime}(z)}{k_BT_0}\right)^2\right]e^{-u(z)/k_BT_0}.
  \nonumber
\end{eqnarray}
Attempting a solution of the form\\
$c_1=c_0(T^{\prime}x/T_0)xf(z)e^{-u(z)/k_BT_0}$ in (\ref{eq:eqc1}),
yields the following equation for the unknown function $f(z)$
\begin{equation}
  f^{\prime\prime}-f^{\prime}\frac{u^{\prime}}{k_BT_0}
  =\frac{u^{\prime\prime}}{k_BT_0}-\left(\frac{u^{\prime}}{k_BT_0}\right)^2
\end{equation}
The solution of this equation is $f=u/k_BT_0+A$, and from the
boundary condition (\ref{eq:bc}) we find that $A=-1$, such that (to
linear order in $T^{\prime}$)
\begin{equation}
  c\simeq c_0\left[1-\frac{T^{\prime}x}{T_0}\left(1-\frac{u(z)}{k_BT_0}\right)
    \right]e^{-u(z)/k_BT_0}
  \label{eq:solconcentration}
\end{equation}
Substituting this expression for the solute concentration in
(\ref{eq:solflux}) yields the solute flux. We find
that the $z$-component is quadratically small in the temperature
gradient: $\vec{j}_n\cdot\hat{z}\sim (T^{\prime})^2$. For the $x$
component, we have that the
\begin{equation}
  j_n(z)=-\left(\frac{T^{\prime}}{T_0}\right)\mu_0c_0u(z)e^{-u(z)/k_BT_0}\hat{x}
    +{\cal O}\left[\left(T^{\prime}\right)^2\right].
    \label{eq:solflux2}
\end{equation}
The total solute flux (per unit length) is obtained by integration,
$j_n= \int_0^{\infty}j_n(z)dz$, and by comparing with
Eq.~(\ref{eq:linmatrix}), we arrive at
    \begin{equation}
  L_{nh}=- \mu_0c_0\int_0^{\infty} u(z)e^{-u(z)/k_BT_0}dz,
\label{eq:lph2}
\end{equation}
which is equal to $L_{hn}$ in Eq.~(\ref{eq:lhp2}).

\section{One-component simple fluids}
\label{sec:1component}

The identical expressions (\ref{eq:lhp2}) and (\ref{eq:lph2}), derived
in the previous section for the transport cross-coefficients, vanish
when the concentration of solute $c_0=0$. This, however, does not mean
that thermo-osmosis does not exist in pure liquids. We recall that
within the model considered above, the solvent was treated as a
structureless medium in which the solute particles are
suspended. There is no wonder, therefore, the limit $c_0\rightarrow 0$
yields no thermo-osmotic effect. In order to study thermo-osmosis in a
pure fluid we must takes its molecular structure into account, as done
in of following section.

Let us consider a setup similar to the one in section
\ref{sec:solsol}, namely a semi-infinite system with a surface at
$z=0$, but this time containing a simple fluid consisting of a single
type of spherical particle. As in the previous section, we denote the
(short-range) interaction between the particles and the surface by
$u(z)$. Let us denote the bulk density of the particles (at
$z\rightarrow\infty$) by $\rho_0$, and in what follows we will neglect
the thermal expansivity of the fluid, i.e., assume that $\rho_0$ is
independent of $T$ within the relevant temperature range.  The latter
assumption is an approximation which is essential for allowing a
simple derivation of analytical expressions for the Onsager cross
coefficients. At a given thermodynamic state ($T(x)$ and $P(x)$,
within a thin slab), the excess surface enthalpy is the change in the
system energy arising from the presence of the surface. This includes
two contributions: (i) direct interactions between the particles and
the surface which are represented by the potential energy $u(z)$, and
(ii) the change in the particle-particle interaction energy resulting
from the variations in the local density of the fluid {\em that are
  due to the particles-surface interactions.}\/ The latter
contribution must be calculated in a statistical manner, i.e. by
averaging over all the particles configurations with the Boltzmann
statistical weights of the isobaric-isothermal ensemble. Taken
together, the ``effective'' surface energy per particle can be
represented by a potential of mean force (PMF), $\phi(z)$, which is
related to the local density by
\begin{equation}
  \rho(z)=\rho_0e^{-\phi(z)/k_BT(x)}.
  \label{eq:pmf}
\end{equation}
The last equation (\ref{eq:pmf}) invokes the LTE approximation (see
discussion in section \ref{sec:solsol}).

A subtle, yet important, point to mention is that in principle the PMF
depends on the local density and temperature, but the approximation
that the fluid has no thermal expansion implies a
temperature-independent PMF. This can be understood from the following
argument: The lack of thermal expansivity means that the density of
the bulk fluid is insensitive to the temperature.  This is an ideal
situation in which a small change in the temperature causes changes in
the intramolecular interactions energy and entropy that exactly cancel
each other. The fact that the fluid has no thermal expansivity does
not mean that the density is uniform throughout the {\em entire
  system}\/ (as, otherwise, there will be no thermo-osmotic
effect). Variations in the local density are encountered at the
interfacial layer as a result of surface interactions (direct and
indirect) that are fully accounted for by Eq.~(\ref{eq:pmf}) that
defines the PMF.  This form represents a mechanical balance between
the average mechanical forces experienced by particles within a thin
layer of width $dz$ ($z<z^{\prime}<z+dz$) and the thermal forces
associated with the mixing entropy of the non-uniform distribution in
the layer. In the absence of thermal expansivity, no other thermal
effects need to be considered, which means that $\phi(z)$ (just like
$\rho_0$), has no temperature dependence.

With the above considerations in mind, we now wish to calculate the
thermo-osmotic and the mechano-caloric Onsager coefficients, and
demonstrate their equivalence. The former is calculated by assuming
that the temperature along the $x$ direction (parallel to surface) is
given by $T(x)=T_0+T^{\prime}x$, and that the pressure is uniform.
Similarly to the case studied in section~\ref{sec:solsol}, the
particle flux in the $z$ directon vanishes to linear order in
$T^{\prime}$. The particle flux per unit volume in the $x$ direction
is given by $\vec{j}_n\cdot\hat{x}=-D(x,z)\partial\rho(x,z)/\partial
x$, where $D(x,z)$ is a coordinate-dependent Fickian diffusion
coefficient.  The $z$-dependence of $D$ arises from the influence of
the surface on the (collective) diffusive dynamics of the
particles. The $x$-dependence, on the other hand, arises from the
temperature gradient and, therefore, is unimportant if one wishes to
calculate the flux $j_n$ {\em to linear order}\/ in $T^{\prime}$. We
thus arrive at the following result:
\begin{eqnarray}
  \vec{j}_n(z)&=&-\hat{x}D(z)\frac{\partial\left[\rho_0
    e^{-\phi(z)/k_BT(x)}\right]}{\partial x}  \label{eq:Jv1}\\
  &=&-\hat{x}\left(\frac{T^{\prime}}{T_0}\right)\rho_0D(z)
  \frac{\phi(z)}{k_BT_0}e^{-\phi(z)/k_BT_0}+ {\cal
    O}\left[\left(T^{\prime}\right)^2\right].\nonumber
\end{eqnarray}
The total particle flux (per unit length in the $y$ direction) is
obtained by integration: $j_n=\int_{0}^{\infty}j_n(z)dz$, and by
comparison with Eq.~(\ref{eq:linmatrix}), we conclude that
\begin{equation}
  L_{nh}=-\int_0^{\infty} \rho_0D(z)\frac{\phi(z)}{k_BT_0}e^{-\phi(z)/k_BT_0}dz.
  \label{eq:lph1}
\end{equation}

Before proceeding to the calculation of the mechano-caloric
coefficient, $L_{hn}$, it is important to dwell shortly on the
difference between the Fickian diffusion coefficient, $D(z)$,
appearing in the previous paragraph and the diffusion coefficient,
$D_0$, which is related to the solute mobility in
section \ref{sec:solsol} via Einstein's relation:
$D_0=\mu_0k_BT_0$. The latter is a single-particle quantity since each
solute particle is treated (in section \ref{sec:solsol}) as an
independent Brownian particle. Single particle diffusion, however, is
not the same as Fickian (collective, or chemical) diffusion, which is
the process relevant to the problem discussed in this section. In
single-component fluids the motion of the particles is strongly
correlated and, clearly, the particles interact with each other via
short-range molecular forces. The confusion between single-particle
and Fickian collective diffusion arises because both processes are
represented by the same diffusion equation. The former describes the
random thermal motion arising from the collisions of a colloidal
particle with the molecules of an embedding fluid medium. The later,
on the other hand, describes the response of a fluid medium to density
inhomogeneities. Of course, in the absence of an external potential,
the fluid would relax to uniform distribution, but this occurs via
random collisions of the fluid particles with each other.

Furthermore, the application of a linear potential (constant force,
$F$) in the fluid results in an exponential Boltzmann equilibrium
distribution if the system is closed. If, one the other hand, the
fluid is found in an open system, the force will cause it to flow with
a velocity field which is linear in the force:
\begin{equation}
  v(z)=\mu(z)F.
  \label{eq:linvf}
\end{equation}
The coefficient of proportionality in this linear velocity-force
relationship (\ref{eq:linvf}), $\mu(z)$, is called Fickian
mobility. As $\mu_z$ is proportional to the velocity field arising
from the application of a constant force like a uniform pressure
gradient, it essentially characterizes the hydrodynamic response of
the system. In the simplest example of a Newtonian Newtonian
incompressible fluid with no-slip boundary conditions, $\mu(z)$ has
the form of a laminar Poiseuille flow.  But Eq.~(\ref{eq:linvf}) is
not limited to this example only, but is a far more general relation
that can be assumed whenever the driving force $F$ is sufficiently
weak (linear response). It is, therefore, also relevant to other
boundary conditions (e.g., slip) and, furthermore, not even limited to
the NS continuum description but can be used in molecular
models. These models are essential for understanding osmosis in
systems where the range of the interaction interfacial regime is of
order of several molecular layer \cite{ganti17,fu17}. Molecular
simulations take into account the variations in the fluid viscosity at
the interfacial layer, which are typically ignored when solving the NS
equation.

The key point to note now is the fact that $\mu(z)$ is related to the
Fickian diffusion coefficient, $D(z)$, via the Nernst-Einstein
equation,
\begin{equation}
  D(z)=\mu(z)k_BT_0.
  \label{eq:nernstein}
\end{equation}
The fact that this relationship takes a similar form to the Einstein
relation for single-particle diffusion is not surprising. It follows
from the fact that the governing diffusion equations in both cases
have the same form (see discussion and derivation in
ref. \cite{mehrerbook}, section 11.3). Equation (\ref{eq:nernstein})
allows us to continue continue with the calculation of the
mechano-caloric coefficient in a fashion similar to the derivation of
its counterpart in the solvent-solute case, Eq.~(\ref{eq:lhp2}). Thus,
we consider a system with constant temperature $T(x)=T_0$ and a small
uniform pressure gradient along the $x$ direction:
$P(x)=P_0+P^{\prime}x$. The fluid density, $\rho(z)$, is given by
Eq.~(\ref{eq:pmf}). The pressure gradient causes a steady-state flow
of the fluid, with a local ($z$-dependent) flux given by
$\vec{j}_n(z)=-\mu(z)P^{\prime}\hat{x}$.  The excess heat flow is
associated with the interfacial free energy that the moving particles
carry which, by definition, is characterizes by the PMF,
$\phi(z)$. Thus, the excess heat (enthalpy) density flux is
$\vec{j}_h(z)=-\mu(z)\rho(z)\phi(z)P^{\prime}\hat{x}$, and by using
the Nernst-Einstein relation this result can be also written as
\begin{equation}
  \vec{j}_h(z)=-\hat{x}P^{\prime}\rho_0D(z)
  \frac{\phi(z)}{k_BT_0}e^{-\phi(z)/k_BT_0}.
  \label{eq:Jq1}
\end{equation}
The heat density flux per unit length in the $y$ direction is
$j_h=\int_0^{\infty}J_h(z)dz$, and from Eq.~(\ref{eq:linmatrix}) we
find that
\begin{equation}
  L_{hn}=-\int_0^{\infty}
  \rho_0D(z)\frac{\phi(z)}{k_BT_0}e^{-\phi(z)/k_BT_0}dz,
\label{eq:lhp1}
\end{equation}
which is identical to $L_{nh}$ in Eq.~(\ref{eq:lph1}).

\section{Discussion and conclusions}
\label{sec:summary}

Derjaguin theory of thermo-osmosis (material flow resulting from a
temperature gradient) is based on Onsager's reciprocity principle
relating thermosmotic flows to the mechno-caloric effect (heat flux
due to a pressure gradient). It is formulated in macroscopic
non-equilibrium thermodynamics terms, but its interpretation at the
molecular level in not entirely clear. Here, we took such an approach
and used statistical-mechanical tools to directly verify the
equivalence between the thermo-osmotic and mechno-caloric transport
coefficients. We considered two very distinct models: (i) an
incopressible (structerless) solvent containing non-interacting solute
particles, and (ii) a single-component fluid without thermal
expansivity. While the calculation of the thermo-osmotic coefficient,
$L_{nh}$, was somewhat complicated, the derivation of the
mechano-caloric coefficient, $L_{hn}$, was simple and
straightforward. In both cases, it followed the same simple logic
recognizing that the excess interfacial heat transport is nothing but
the product of the volume material flux arising from the application
of a pressure gradient force and the surface interaction free energy
of the moving particles. The latter can be associated with the
potential of mean force of the surface, which is directly related to
the equilibrium distribution of the the solute and the fluid particles
(case model ii). The only difference between expressions
(\ref{eq:lhp2}) and (\ref{eq:lhp1}) for the cross-coefficients in both
cases, is the particle mobility relevant to the situation. In the
solvent-solute model, we assumed a constant single particle mobility,
$\mu_0$, since each solute particle was treated as an independent
Brownian particle. We could, in fact, correct this result to account
for the particle's hydrophobic interactions with the surface, by using
Brenner's theory \cite{brenner61}. In contrast, the mobility in the
single-component fluid model, is dominated by the particle-particle
interactions and, thus, one has to consider the collective Fickian
mobility $\mu(z)=D(z)/k_BT_0$ that, in principle, depends on the fluid
density. Thus, the major difficulty in evaluating the transport cross
coefficients in simple fluids is to determine the mobility as a
function of the distance from the surface. Derjaguin circumvented this
problem by considering the hydrodynamic flow profile rather than the
local mobility, but this continuum picture is relevant only when the
range of the surface potential energy is much larger than several
molecular layers.

We conclude by considering yet another classical example, which is a
1:1 electrolyte such as sodium chloride interacting with a uniformly
charged surface.  This problem resembles the solvent-solute model
discussed in the this paper, with two important differences: First, in
the electrolyte problem we have two types of particles (cations and
anions) suspended in the solvent (water), rather than one. Second, the
ions have Coulomb interactions not only with the charged surface but
also with each other. The equilibrium distribution of the ions can be
found from the solution of the Poisson-Boltzmann equation. In the
Debye-H\"uckel approximation, the electrostatic potential is
exponentially screened by the ions in the diffusive electric double
layer \cite{andelman}
\begin{equation}
  \psi(z)=\psi_se^{-z/\lambda_D},
  \label{eq:dhphi}
\end{equation}
where $\psi_s$ is the surface potential, and
\begin{equation}
  \lambda_D=\sqrt{\frac{\epsilon k_BT}{2c_0e^2}}
    \label{eq:screening}
\end{equation}
is the Debye screening length, where $\epsilon$ is the water
permitivity ($\epsilon\simeq 78\epsilon_0$), and $e$ is the electron
charge. The an- and cations distributions, $c_-$ and $c_+$
respectively, are related to $\psi(z)$ via relations similar to
Eq.~(\ref{eq:pmf2}):
\begin{equation}
  c_{\mp}(z)=c_0e^{\pm e\psi(z)/k_BT}.
  \label{eq:pmf3}
\end{equation}
The excess electrostatic free energy density is given by
\begin{eqnarray}
  f_{\rm el}(z)&=&\frac{\epsilon}{2}\left(\frac{d\psi}{dz}\right)^2+
  k_BT\sum_{i=\pm} \left[c_i\ln\left(\frac{c_i}{c_0}\right)-c_i+c_0\right]
  \nonumber \\
  &=&\frac{\epsilon\psi_s^2e^{-2z/\lambda_D}}{\lambda_D^2}=
  2c_0\frac{(e\psi_s)^2}{k_BT}e^{-2z/\lambda_D}.
  \label{eq:felectro}
\end{eqnarray}
In the Debye-H\"uckel limit ($e\psi_s\ll k_BT$), this electrostatic
free energy density can be equally divided between the an- and
cations. To leading order, we can assume that their densities in the
boundary layer are nearly identical and, thus, the relevant mobility for
the transportation of $f_{\rm el}$ under the action of a constant
pressure gradient is $(\mu_-+\mu_+)/2$. The mechano-caloric
coefficient is thus given by
\begin{eqnarray}
  L_{hn}&=& \frac{\mu_-+\mu_+}{2}\int_0^{\infty}f(z)dz
  \nonumber \\
  &=&\frac{\mu_-+\mu_+}{2}c_0\lambda_D
  \frac{(e\psi_s)^2}{k_BT}.
\label{eq:lhp3}
\end{eqnarray}
Notice the unusual dependence on the bulk concentration $L_{hn}\sim
c_0\lambda_D\sim c_0^{1/2}$, which is an manifestation of the
collective nature of the screening effect. This is a major difference
from the example of non-interacting solute particles discussed in
section \ref{sec:solsol}, which also complicates the calculation of
the thermo-osmotic coefficient $L_{nh}$. Further complications in the
analysis of the thermo-osmotic effect in electrolytes arise from the
temperature-dependence of the permitivity of the
water \cite{fayolle08,rasuli08}.

\section{Acknowledgments}

I thank Daan Frenkel for numerous insightful discussions on the topic
and for comments on the manuscript. This work was supported by the
Israel Science Foundation (ISF) through Grant No. 991/17.

\section{Authors contributions}

OF conducted the research and wrote the manuscript.

\appendix
\section{Hydrodynamic considerations}
\label{sec:hydro}

In section~\ref{sec:solsol} we considered the model of an
incompressible liquid containing a small concentration of solute
molecules. We derived identical expressions (\ref{eq:lhp2}) and
(\ref{eq:lph2}) for the cross-coefficients, relating them to
equilibrium properties of the solute.  The only quantity in these
expressions that depends on properties of the solvent is the mobility
of the solute-particles which, presumably, also depends on the
viscosity, $\eta$, of the embedding medium. As already discussed
above, Onsager reciprocity establishes a relationship between the
mechno-caloric and thermo-osmotic flows, but it does not provide
sufficient information from which the thermo-osmotic force on the
surface and the slip-velocity can be derived. In order to calculate
the latter, one must consider the solvent flow, which is done by
invoking the NS equation for the flow of an incopmressible fluid with
constant viscosity $\eta$:
\begin{equation}
  \eta \nabla^2\vec{v}=\vec{\nabla}\Pi(x,z)-\vec{f}_{\rm body},
  \label{eq:navierstokes}
\end{equation}
where $v$ is the solvent velocity, $f_{\rm body}$ is the body force
exerted on a volume element by the solute-surface interactions, and
$\Pi$ is the fluid hydrodynamic pressure, which is not identical to
the solute osmotic pressure that enters into the above calculation of
Onsager cross-coefficients.

For the completeness of the discussion, we will calculate the slip
velocity here by following the hydrodynamic calculation of
Anderson~\cite{anderson82} for the diffuso-osmotic slip. The basic
assumption is that, within the boundary layer, the normal velocity,
$v_z$, is negligible. Therefore, for the $z$ component the NS equation
reads
\begin{eqnarray}
  \frac{\partial \Pi}{\partial z}&=&f_{\rm body}^z=-c(x,z)u^{\prime}(z)
    \label{eq:ns1}\\
  &=&-u^{\prime}(z)c_0\left[1-\frac{T^{\prime}x}{T_0}
    \left(1-\frac{u(z)}{k_BT_0}\right)
    \right]e^{-u(z)/k_BT_0},\nonumber
\end{eqnarray}
where Eq.~(\ref{eq:solconcentration}) for the solute concentration has
been used in the last equality of Eq.~(\ref{eq:ns1}). This non-linear
equation has no simple analytic solution, except in the limit of a
weak interaction potential $u(z)/k_BT_0\ll 1$, in which case the
equation simplifies to\\
$\partial_z\Pi=k_BT_0C_0(1-T^{\prime}x/T_0)\partial_z[\exp(-u/k_BT_0)]$
and, thus,
\begin{equation}
  \Pi(x,z)=k_BT_0c_0\left[1-\frac{T^{\prime}x}{T_0}\right]e^{-u(z)/k_BT_0}
  +\Pi_0(x).
  \label{eq:ns2}
\end{equation}
From this result, we identify $\Pi_0(x)$ as the difference between the
hydrodynamic and osmotic pressures. This component can be found by
invoking the requirement that the bulk hydrodynamic pressure away from
the surface becomes uniform. Thus,
$\Pi_0(x)=-k_BT_0c_0(1-T^{\prime}x/T_0)+\Pi_0$, and the hydrodynamic
pressure is given by
\begin{equation}
  \Pi(x,z)=k_BT_0c_0\left(1-\frac{T^{\prime}x}{T_0}\right)
  \left[e^{-u(z)/k_BT_0}-1\right]+\Pi_0,
  \label{eq:ns3}
\end{equation}
where $\Pi_0$ is an unimportant constant. 

In order to find the interfacial shear stress, we return to the NS
equation for the $x$ component. Here, we assume that $v_x$ varies much
more rapidly along the normal $z$ direction than parallel to the
surface in the $x$ direction. Thus, $\nabla^2 v_x\simeq \partial^2
v_x/\partial z^2$, and considering that $f_{\rm body}^x =0$ (since the
surface forces on the solute particles are in the $z$ direction), the
NS equation for $v_x$ (\ref{eq:navierstokes}) reads
\begin{equation}
  \eta\frac{\partial^2 v_x}{\partial z^2}=\frac{\partial \Pi}{\partial x}=-
  -k_BT^{\prime}c_0\left(e^{-u(z)/k_BT_0}-1\right).
  \label{eq:ns4}  
\end{equation}
The shear stress, $\sigma_{xy}$, is then found by integrating
Eq.~(\ref{eq:ns4}) with respect to $z$
\begin{equation}
  \sigma_{xz}(z)=\eta\frac{\partial v_x}{\partial z}=k_BT^{\prime}c_0
  \int_z^{\infty}\left(e^{-u(z^{\prime})/k_BT_0}-1\right)dz^{\prime},
  \label{eq:shear1}
\end{equation}
where the integration constant was determined such that
$\sigma_{xz}(z\rightarrow\infty)=0$. The thermo-osmotic force per unit
surface area, $\sigma_{\rm t-o}$, is nothing but the shear stress
evaluated at $z=0$:
\begin{equation}
  \sigma_{\rm t-o}=k_BT^{\prime}c_0
  \int_0^{\infty}\left(e^{-u(z)/k_BT_0}-1\right)dz.
  \label{eq:shear2}
\end{equation}
Remarkably, it is expressed as a function of only equilibrium
properties of the solute, and has no dependence on the solvent
viscosity or the type of boundary conditions (slip/stick) assumed for
the fluid flow. The fluid velocity profile , $v_x(z)$, can be obtained
by a second integration of Eq.~(\ref{eq:ns4}):
\begin{equation}
  v_x(z)=\frac{k_BT^{\prime}c_0}{\eta} \int_0^zdz^{\prime}
  \int_{z^{\prime}}^{\infty}\left(e^{-u(z^{\prime\prime})/k_BT_0}-1\right)
  dz^{\prime\prime},
  \label{eq:shear3}
\end{equation}
and the integration constant, in this case, was set to satisfy no-slip
boundary conditions: $v_x(z=0)=0$.


\begin{thebibliography}{}

\bibitem{marbach19} S. Marbach, L. Bocquet, Chem. Soc. Rev.,
  {\bf 48}, 3102 (2019).

\bibitem{wall10} S. Wall, Curr. Opin. Colloid Interface Sci. {\bf 15},
  119 (2010).
  
\bibitem{bacchin19} P. Bacchin, K. Glavatskiy, V. Gerbaud,
  Phys. chem. Chem. Phys. {\bf 21}, 10114 (2019).

\bibitem{barragan17} V. M. Barrag\'an, S. Kjelstrup,
  J. Non-equilib. Thermodyn. {\bf 42}, 217 (2017).

\bibitem{derjaguinbook} B. Derjaguin, N. Churaev, V. Muller, {\em
  Surface Forces}\/ (Plenum, New York, 1987).

\bibitem{stone04} H. A. Stone, A. D. Stroock, A. Ajdari,
  Annu. Rev. Fluid Mech. {\bf 36}, 381 (2004).

\bibitem{imbrongo16} J. Imbrogno, G. Belfort,
  Annu. Rev. Chem. Biomol. Eng. {\bf 7}, 29 (2016).

\bibitem{biologybook} H. Lodish {\em et al.}, {\em Molecular Cell
  Biology}\/ 4th ed. (W. H. Freeman, New York, 2000).

\bibitem{lippmann1907} G. Lippmann, C. R. Acad. Sci. {\bf 145}, 104
  (1907).

\bibitem{onsager1} L. Onsager, Phys. Rev. {\bf 37}, 405 (1931).

\bibitem{onsager2} L. Onsager, Phys. Rev. {\bf 38}, 2265 (1931).

\bibitem{bergulla16} A. P. Bregulla {\em et al.}\/,
  Phys. Rev. Lett. {\bf 116}, 188303 (2016).

\bibitem{ganti17} R. Ganti, Y. Liu, D. Frenkel, Phys. Rev. Lett. {\bf
  119}, 038002 (2017).

\bibitem{grootmazurbook} S. R. De Groot, P. Mazur, {\em
  Non-Equilibrium Thermodynamics}\/ (North-Holland, Amsterdam, 1969).

\bibitem{anzini19} P. Anzini, G. M. Colombo, Z. Filiberti, A. Parola,
  Phys. Rev. Lett. {\bf 123}, 028002 (2019).

\bibitem{wurger10} A. W\"urger, Rep. Prog. Phys. {\bf73}, 126601
  (2010).

\bibitem{berreta04} G. P. Beretta, E. P. Gyftopoulos,
  J. Chem. Phys. {\bf 121}, 2718 (2004).

  
\bibitem{anderson82} J. L. Anderson, M. E. Lowell, D. C. Prieve, J
  . Fluid Mech. {\bf 117}, 107 (1982).

\bibitem{fayolle08} S. Fayolle, T. Bickel, A. W\"urger, Phys. Rev. E
  {\bf 77}, 041404 (2008).

\bibitem{farago19} O. Farago, Phys. Rev. E {\bf 99}, 062108 (2019).

\bibitem{piazza08} R. Piazza, A. Parola, J. Phys. Condens. Matter {\bf
  20}, 153102 (2008).

\bibitem{brenner61} H. Brenner, Chem. Eng. Sci. {\bf 16}, 242 (1961).

\bibitem{mehrerbook} H. Mehrer, {\em Diffusion in Solids:
  Fundamentals, Methods, Materials, Diffusion-Controlled Processes}\/
  (Springer, Berlin, 2007).

\bibitem{fu17} L. Fu, S. Merabia, L. Joly, Phys. Rev. Lett. {\bf 119},
  214501 (2017).

\bibitem{andelman} D. Andelman, Chap. 12 Vol. 1 in {\em Handbook of
  Biological Physics}\/, edited by R. Lipowsky and E. Sackmann
  (Elsevier Amsterdam, 1995).

\bibitem{rasuli08} S. N. Rasuli, R. Golestanian, Phys. Rev. Lett. {\bf
  101}, 108301 (2008).

  
\end{thebibliography}
\end{document}